\documentclass[10pt,prb,aps,superscriptaddress,amsmath,amssymb,twocolumn,floatfix,longbibliography]{revtex4-1}
\usepackage{mathtools}
\usepackage{mathrsfs, amsmath, wasysym}
\usepackage{bbold}
\usepackage{dsfont}
\usepackage{braket,mleftright}
\usepackage[mathscr]{eucal}
\usepackage{xfrac}
\usepackage{graphicx}
\usepackage{dcolumn}
\usepackage{bm}
\usepackage{manfnt}
\usepackage{color}
\usepackage{tabularx}
\usepackage{natbib}
\usepackage[bookmarksopen, colorlinks,linkcolor=blue,citecolor=blue,urlcolor=blue]{hyperref}
\usepackage[normalem]{ulem}
\usepackage[all]{hypcap}
\usepackage[dvipsnames,usenames,table]{xcolor}

\newcommand{\bg}{\begin{gather}}
\newcommand{\eg}{\end{gather}}

\newcommand{\be}{\begin{equation} }
\newcommand{\ee}{\end{equation}}
\newcommand{\beq}{\begin{eqnarray}}
\newcommand{\eeq}{\end{eqnarray}}

\newcommand{\bk}{{\bf k}}

\newcommand{\bA}{{\bf A}}

\newcommand{\br}{{\bf r}}

\newcommand{\bG}{{\bf G}}

\newcommand{\Hbdg}{\mathcal{H}_{\text{BdG}}}

\newcommand{\R}{\mathcal{R}}

\begin{document}

\title{Unified topological phase diagram of quantum Hall and superconducting vortex-lattice states}

\author{Daniil S. Antonenko}
\email{daniil.antonenko@yale.edu}
\affiliation{Department of Physics, Yale University, New Haven, Connecticut 06520, USA}
\author{Liang Fu}
\affiliation{Department of Physics, MIT, Cambridge, USA}
\author{Leonid I. Glazman}
\affiliation{Department of Physics, Yale University, New Haven, Connecticut 06520, USA}
\date{\today}

\begin{abstract}

We present the global topological phase diagram of a two-dimensional electron gas placed in a quantizing magnetic field and proximitized by a superconducting vortex lattice. Our theory allows for arbitrary ratios of the pairing amplitude, magnetic field, and chemical potential. 
By analyzing the Bogoliubov–de Gennes Hamiltonian, we show that the resulting phase diagram is highly nontrivial, featuring a plethora of topological superconducting phases with chiral edge modes of quasiparticles. 
Landau-level mixing plays an essential role in our theory: even in the weak-pairing limit, it generically splits the integer quantum Hall transition lines into a sequence of transitions with larger Chern number jumps of both signs protected by the symmetries of the superconducting vortex lattice.
Interestingly, we find that weak pairing induces trivial or topological superconductivity when chemical potential is tuned to a Landau level energy, depending on the Landau level index.

\end{abstract}

\maketitle

\section{Introduction}
The recent progress in proximitizing a quantum well formed in a semiconductor by a superconductor epitaxially grown on top of it has renewed interest in the interplay of superconductivity with the quantum Hall effect \cite{MakingSwaveTopological,  McDonald_new,  MishmashYazdaniZaletel, Jain2022helical}. The electron states in a superconductor and in a two-dimensional electron gas subject to a quantizing magnetic field are topologically different. 
Even the hybridization of the two simplest phases, the $s$-wave superconductivity and the integer quantum Hall effect, is far from trivial. The topology of electrons occupying an integer number $N$ of Landau levels manifests itself in the presence of $N$ gapless states propagating along the edges of a quantum Hall system. The nontrivial topological state is protected by a finite cyclotron gap $\hbar\omega_c$ and remains intact as long as the pairing potential $\Delta(\mathbf r)$ induced by a superconductor remains weak, $|\Delta(\mathbf r)|\ll \hbar\omega_c$. In the opposite limit of a weak magnetic field, $|\Delta(\mathbf r)|\gg \hbar\omega_c$, one expects formation of topologically-trivial superconducting $s$-wave state. The spatial structure of the complex-valued function $\Delta(\mathbf r)$ is determined by the lattice of vortices induced by the magnetic field in the superconductor\cite{Abrikosov_vortex_lattice}. The spatially nonuniform pairing potential $\Delta(\mathbf r)$ along with the magnetic field induces in the proximitized semiconductor narrow quasiparticle bands \cite{Franz-Tesanovic, Liu-Franz} of hybridized Caroli--de Gennes--Maricon levels\cite{CdGM}, separated from the Fermi level by a small but finite gap. This quasiparticle gap protects the topologically-trivial state, which has no gapless edge modes.

In our previous work\cite{MakingSwaveTopological}, we showed that introducing superconducting correlations into quantum Hall state while keeping the integer number of occupied levels $N$ fixed, leads to a sequence of topological phase transitions. 
As the superconducting order parameter $\Delta$ (either proximity-induced or intrinsic) increases, the superconducting Chern number undergoes multiple jumps and eventually drops to zero when the system evolves into the conventional Abrikosov-lattice state of an $s$-wave superconductor in a magnetic field. Remarkably, we found that the Chern number can change by relatively large integers, up to $12$ across a single transition. In those calculations, the chemical potential $\mu$ was fixed at the midgap between Landau levels of the unperturbed quantum Hall state, $\mu = E_N + \hbar\omega_c /2$. 
\begin{figure}[h!]
\includegraphics[width=0.45\textwidth]{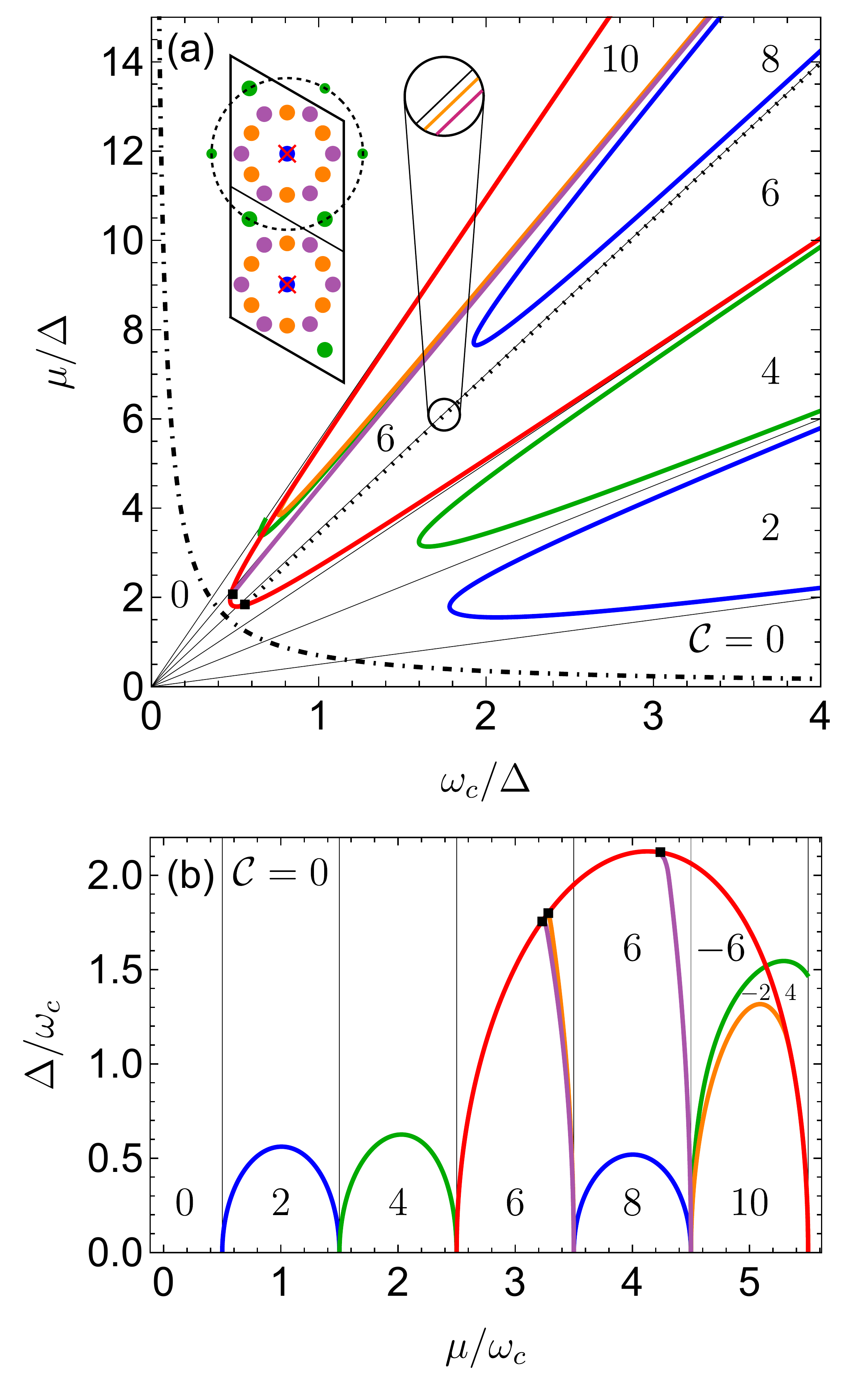}
	\caption{
    (Triangular vortex lattice) Topological phase diagram of the two-dimensional electron gas in the presence of magnetic field and superconducting order parameter with Abrikosov vortex lattice in two different axes choice. Here $\omega_c$ is the cyclotron frequency, $\mu$ is the chemical potential and $\Delta$ is the superconducting order parameter amplitude. The dashed line ($\omega_c \mu \sim \Delta^2$) is an estimate for the boundary of the region containing topological domes, see Eq.~\eqref{top-trivial-boundary}. The dotted line shows a pair of closely lying transition lines, illustrated in the zoom-in. The black squares mark the tricritical points. Inset: configuration of the degenerate gap closure points in the magnetic Brillouin zone; filled circles mark Dirac points while crosses stand for the cubic band touchings. The colors correspond to the colors of the lines in the main plot. 
    As shown in Sec.~\ref{sec:symmetry}, blue points and red crosses coincide with the $C_6$ rotation axes; the other groups of same-color points are symmetric under this operation. When moving along the transition lines in the main plot, the green, orange, and purple points move in the radial direction retaining the $C_6$ symmetry.
	\label{fig:tr-phase-diagram}
    }
\end{figure}

By contrast, several other works\cite{DukanTesanovic, MishmashYazdaniZaletel, ZocherRosenow, Nomura-pwave-proximity}
considered a complementary limit: a single Landau level perturbed by the superconducting vortex-lattice pairing potential $\Delta(\mathbf{r})$. This approximation is supposed to be valid when the matrix elements of $\Delta(\mathbf{r})$ are much smaller than the Landau-level spacing $\hbar \omega_c$, and when the chemical potential lies close to the chosen Landau level. 
In that regime, the vortex-lattice order parameter $\Delta(\mathbf{r})$ induces a quasiparticle band dispersion featuring multiple gap-closing points when the chemical potential is tuned to the Landau-level energy, $\mu = E_N$. For a spin-degenerate electron gas without Zeeman or spin-orbit coupling, proximitized by an $s$-wave superconductor, the Chern number changes by two as the chemical potential crosses a Landau level. 

The large Chern number jumps reported in Ref.~\onlinecite{MakingSwaveTopological}  versus the $\Delta \mathcal{C} = 2$ transitions found in the single-Landau-level picture suggest that the phase diagram of superconductivity and quantum Hall effect as a function of density and magnetic field is richer than previously known and calls for a complete understanding. This motivates the present study. 
In this work, we construct the complete topological phase diagram of a two-dimensional electron gas proximitized by a superconducting vortex lattice. For the triangular vortex lattice, the result is shown in Fig.~\ref{fig:tr-phase-diagram}. Similar to Ref.~\onlinecite{MakingSwaveTopological}, we do not observe odd Chern numbers in our model, so Majorana edge modes do not appear. Instead, at moderate values of $\Delta/\omega_c$, we predict 
the presence of topologically protected edge modes of
Dirac fermions which can be probed through transport and tunneling measurements.

Finally, we note that our theory applies equally well to proximity-induced and intrinsic superconductivity. While the former route is currently more accessible experimentally\cite{Manfra_InAs_MobilityExceeding}, we expect that the coexistence of intrinsic superconductivity and quantum Hall quantization can be realized in several two-dimensional materials, such as graphene multilayers\cite{ParkNature}.

We work at zero temperature ($T = 0$) and take $\hbar = c = 1$ throughout the manuscript.

\section{Model}
\label{sec:model}

We consider a two-dimensional electron gas with a parabolic dispersion subject to the out-of-plane magnetic field $B$, which is assumed to be uniform even in the presence of the superconducting correlations; the latter point is valid since the penetration length diverges in the two-dimensional limit. In the absence of superconductivity, electrons are described by the Hamiltonian 
\be \label{H0}
H_0 = (-i\bm{\nabla} - e \bA/c)^2 / 2m - \mu,
\ee
where $\bA = (-By, 0, 0)$ is magnetic vector potential in the Landau gauge. We neglect the Zeeman splitting assuming a small electron effective mass. $N$-th Landau level has the energy $E_N = (N + 1 / 2) \omega_c - \mu$, where $\omega_c = e B / (m c)$ is the cyclotron frequency. Hamiltonian \eqref{H0} defines a continuum model, which serves as our starting point. Therefore, we do not introduce an atomic lattice, and the issue of commensurability between atomic and superconducting vortex lattices\cite{Hofstadter, Harper, Jain2022helical, Shaffer-HofstadterSC} is irrelevant for our analysis.

\begin{figure}[h]
	\includegraphics[width=0.48\textwidth]{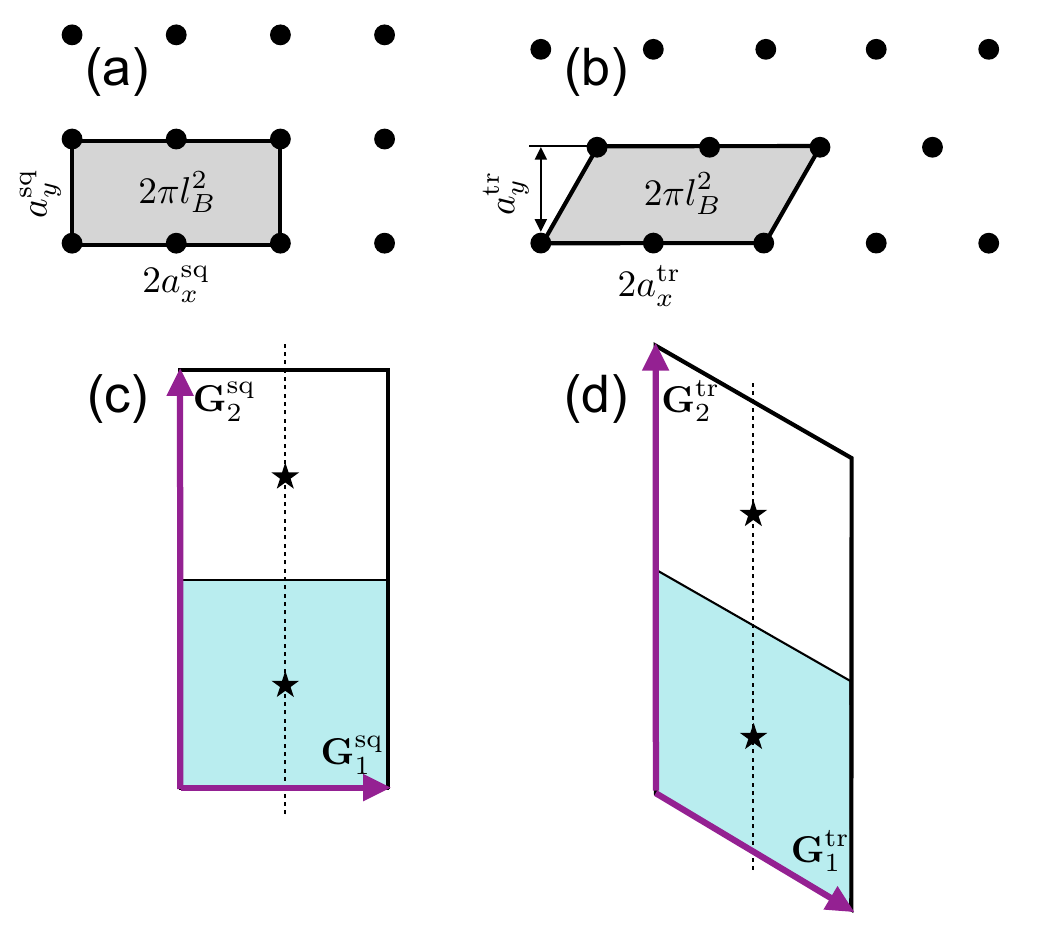}
	\caption{(a, b) Superconducting vortex lattices with the (a) square and (b) triangular configurations. Magnetic unit cell have the area $2\pi l_B^2$ in both cases and is highlighted in grey.
    (c, d) Magnetic Brillouin zones for the (c) square vortex lattice and (d) triangular vortex lattice; reciprocal lattice vectors $\bG_1$ and $\bG_2$ are shown in purple. A half of each BZ is highlighted in turquoise; the dispersion in the other half is identical due to the symmetry \eqref{half-magnetic-transform}. Stars represent the centers of the inversion and rotation symmetries ($C_4$ and $C_6$ correspondingly); dashed line depicts one of the mirror symmetries of the dispersion.
	\label{fig:Lattices}}
\end{figure}

We introduce superconducting correlations to our model through the mean-field Bogoliubov-de Gennes Hamiltonian \cite{Bogoliubov, deGennes_book}:
\be \label{BdG}
	\Hbdg = 
	\begin{pmatrix}
		\hat{c}_{\uparrow} \\ \hat{c}_{\downarrow}^\dagger
	\end{pmatrix}^{\dagger}
	 \begin{pmatrix}
		H_0 & \Delta (\br) \\
		\Delta^*(\br) & -H_0^T
	\end{pmatrix}
	\begin{pmatrix}
		\hat{c}_{\uparrow} \\ \hat{c}_{\downarrow}^\dagger
	\end{pmatrix}.
\ee
Following our previous work\cite{MakingSwaveTopological},we adopt the Abrikosov form of the pairing potential $\Delta(r)$\cite{Abrikosov_vortex_lattice}, which can appear due to the proximity to a bulk superconductor or as an intrinsic phenomenon. Abrikosov form is quantitatively accurate in the vicinity of the phase transition, where Ginzburg-Landau description is valid. Moreover, it captures all symmetries of the vortex lattice and magnetic translation group, so we expect that the outcomes of our calculations and our results on topology remain qualitatively correct in the whole phase diagram. For the purpose of computation, Abrikosov vortex lattice for the square [Fig.~\ref{fig:Lattices}(a)] and triangular [Fig.~\ref{fig:Lattices}(b)] vortex arrangement can be conveniently represented as a series:
\be \label{Delta_r}
	 \Delta(\br) = \Delta \sum_{t=-\infty}^{\infty} C_t
     \Phi_t (\br)
\ee
where $\Phi_t (\br)$ is the lowest Landau level wavefunction for a charge-$2e$ particle with the orbit center at $y_t = 2 t a_y$, where $t$ is an integer.
The coefficients are $C_t = 1$ for the square lattice and $C_t = \exp (-i \pi t^2 / 2)$ for the triangular lattice (despite triangular lattice being the most physically relevant case, we find it instructive to consider the square vortex arrangement as well.) Note that the pairing potential $\Delta(\br)$ is quasiperiodic and satisfies magnetic translation symmetry.

To proceed, we consider the energy spectrum for Bogoliubov quasiparticles in the magnetic Brillouin zone.  Since a Cooper pair carries charge $2e$ and, correspondingly, each vortex hosts a superconducting flux quantum $h/2e$, the Bogoliubov (magnetic) unit cell, highlighted in gray in Fig.~\ref{fig:Lattices}(a,b), is twice as large as the vortex-lattice unit cell\cite{Franz-Tesanovic, Liu-Franz}. Consequently, the magnetic Brillouin zone of quasiparticles is two times smaller, see Fig.~\ref{fig:Lattices}(c,d). Further technical details for the pairing potential \eqref{Delta_r} and the magnetic Bloch wavefunctions can be found in Appendix~\ref{app:magnetic-bloch-states} \cite{DukanTesanovic, TesanovicSacramento, McDonald4}.

\section{Overview}
\label{sec:single-LL}

In this Section, we overview the results of the single–Landau-level approximation in the weak pairing regime $\Delta\ll \omega_c$ \cite{DukanTesanovic, McDonald6, MishmashYazdaniZaletel} and show that it misses an important part of the physics of the problem. Specifically, the topological transitions in this approximation always occur at $\mu = E_N$ and exhibit a Chern number jump of two, shown as slanted thin black lines in Figs.~\ref{fig:tr-phase-diagram}(a) and \ref{fig:sq-phase-diagram}(a). In contrast, as we will show in the next Section, accounting for Landau level mixing makes the topological transition lines split into several separate transition trajectories, even in the limit $\Delta \ll \omega_c$. Globally, the topological phase diagram displays a rich dome-shaped structure, as shown in Figs.~\ref{fig:tr-phase-diagram}(b) and \ref{fig:sq-phase-diagram}(b). 

The single-Landau-level projection is naively justified if the matrix elements of the pairing potential are small, $|\Delta_{NM}(\bm{k})| \ll \omega_c$, and the chemical potential lies close to a Landau level: $|\mu - E_N| \ll \omega_c$. Then, in the leading order of the perturbation theory one finds\cite{DukanTesanovic, McDonald6, MishmashYazdaniZaletel}: 
\be \label{Dispersion-1order}
    \epsilon_N = \sqrt{(\mu - E_N)^2 + |\Delta_{NN}(\bm{k})|^2},
\ee
where $\Delta_{NM}(\mathbf{k})$ are given by Eqs.~\eqref{Delta_NM_k_sq} and \eqref{Delta_NM_k_tr}.
In this approximation, the gap closes when chemical potential exactly matches the energy of a Landau level, $\mu = E_N$, and, in addition, the matrix element $\Delta_{NN}(\bm{k})$ vanishes. We show the correspinding points $\mathbf{k}_i$ in Fig.~\ref{fig:gap-closures} for $N = 0$--$4$. By continuity to the $\Delta = 0$ case, we know that the Chern number changes by $2$ upon $\mu$ crossing the Landau level energy $E_N$ for any $N$; correspondingly, the sum of the topological charges at $\mathbf{k}_i$ for any given $N$ also equals two.

The leading order approximation Eq.~\eqref{Dispersion-1order} suffers from the following issue: not all gap-closing points are related by space group symmetry of the vortex lattice (which we study in detail in Sec.~\ref{sec:symmetry}). In other words, zero-energy $\bk$ points of Eq.~\eqref{Dispersion-1order} have accidental degeneracies. Therefore, the single-Landau-level approximation is inadequate and should be  extended to include Landau level mixing. We perform such extension in Sec.~\ref{sec:many-LLs}, demonstrating that accidental degeneracy is lifted. 
As a result, the topological phase boundary splits into several transition lines that remain close to the Landau level line $\mu = E_N$ at $\Delta \ll \omega_c$, leading to intermediate phases with various Chern numbers. The respective Chern number jumps sum up to $\Delta \mathcal{C} = 2$ as chemical potential sweeps across a Landau level.

\section{Topological phase diagram}
\label{sec:many-LLs}
In this Section we extend our calculation beyond the single-Landau level limit, considered in Section~\ref{sec:single-LL}. After discussing the qualitatively new effects that emerge already in the second order of the perturbation theory, we perform a calculation in a Hilbert space 
that includes many Landau levels.
We follow the calculation procedure outlined in our previous work\cite{MakingSwaveTopological} generalizing it to the case of arbitrary ratio $\mu / \omega_c$ (in the previous work we took the chemical potential exactly in the middle between two Landau levels, $\mu = E_N + \omega_c / 2$, which corresponds to an even filling factor, $\nu = 2N$). We now study the whole phase diagram up to the filling factor $\nu = 10$, corresponding to five filled Landau levels. For each  topological transition, we identify the whole transition trajectory in the $(\omega_c / \Delta$, $\mu / \Delta)$ plane and study in detail the nature of the gap closure points, see Figs.~\ref{fig:tr-phase-diagram} and \ref{fig:sq-phase-diagram}. These results match with the large Chern number jumps observed in Ref.~\onlinecite{MakingSwaveTopological}  and reveal the actual split-transition behavior of the system at $\omega_c / \Delta \gg 1$ proving the insufficiency of the strict single-Landau-level limit.

\subsection{Next-order perturbation theory}
\label{sec:next-order-perturbation}
We start by extending the perturbative (in $\Delta / \omega_c$) expression \eqref{Dispersion-1order} to the next order. As a result, the values $E_N$ in Eq.~(\ref{Dispersion-1order}) acquire a $\bm k$-dependent correction,
\be \label{Dispersion-2order}
    \delta E_N(\bm{k}) = \sum_{M\neq N}\frac{|\Delta_{NM}(\bm{k})|^2}{E_N - E_M},
\ee
due to admixture of Landau levels with $M\neq N$.
A simple asymptotic analysis of Eqs.~\eqref{Delta_NM_k_sq}--\eqref{Delta_NM_k_tr} shows that at fixed $N$, the matrix elements decay exponentially, $\Delta_{NM}(\bm{k}) \sim 2^{-M/2}$ at $M\to\infty$, where the polynomial prefactors were ignored. Such fast decay is due to the rapidly oscillating wavefunction in the convolution \eqref{Delta_Landau_basis}. (Note that the diagonal matrix elements, $\Delta_{NN}(\bm{k})$ decay only as a power law, $N^{-1/4}$, as was shown in Ref.~\onlinecite{MakingSwaveTopological}.) We therefore conclude that the sum \eqref{Dispersion-2order} for low Landau levels converges fast, $|M - N|\sim 1$. 

\begin{figure}\includegraphics[width=0.47\textwidth]{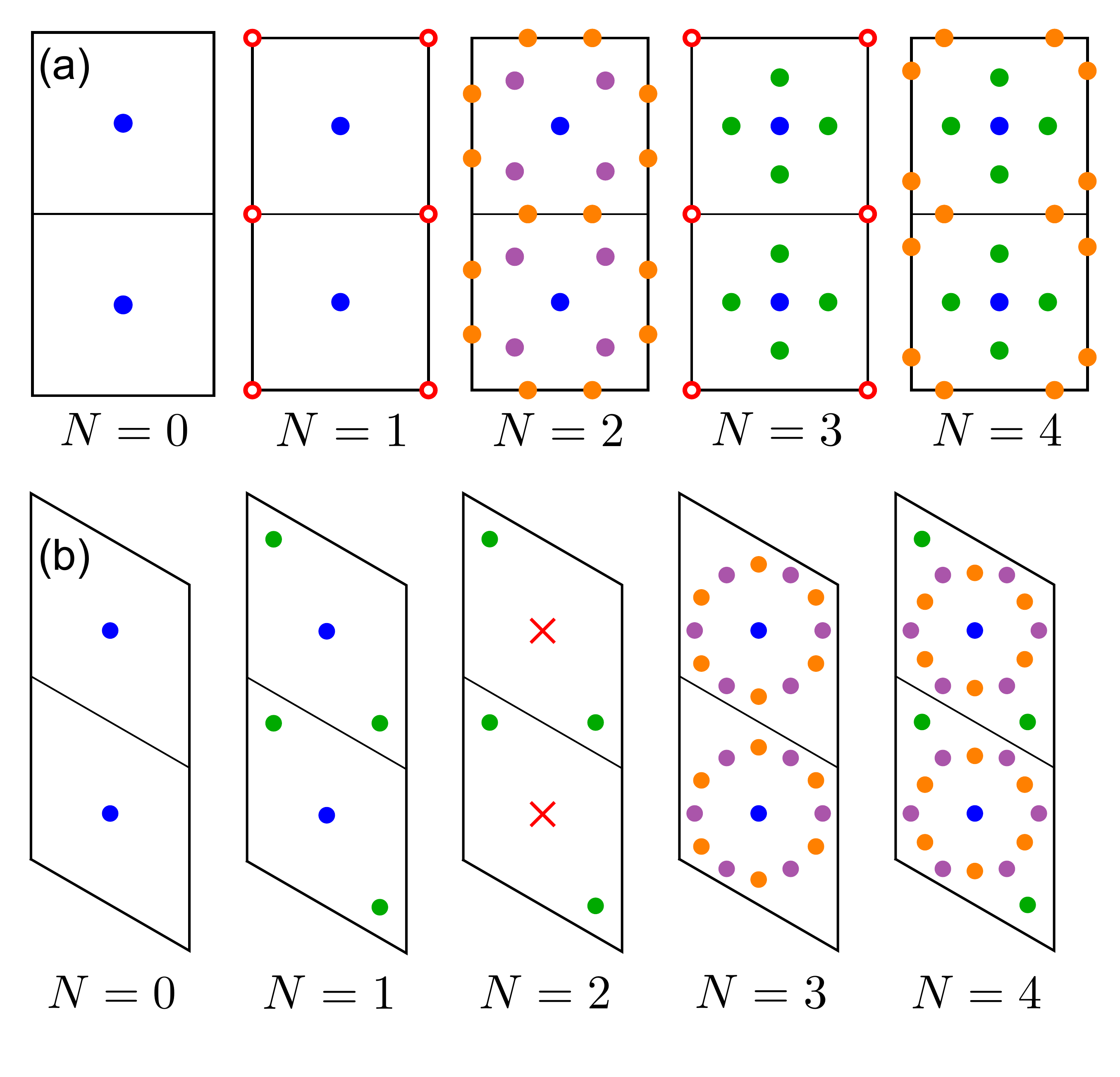}
    \caption{Gap closure points in magnetic Brillouin zone in the single Landau level limit studied in Section~\ref{sec:single-LL} for the (a) square vortex lattice; (b) triangular vortex lattice. This regime requires $\omega_c \gg \Delta$ and chemical potential close to the energy of the $N$-th Landau level, $N = 0$--$4$. Colors of the markers match the colors of the topological phase transition lines of Figs.~\ref{fig:tr-phase-diagram} and \ref{fig:sq-phase-diagram}. Filled circels represent Dirac points, empty circles---quadratic band crossings, and crosses---cubic band crossings.
\label{fig:gap-closures}
}
\end{figure}

The pattern of zeros $\mathbf{k}_i$ of the diagonal matrix element $\Delta_{NN}(\mathbf{k})$ is invariant under the symmetry group of the superconducting vortex lattice; its specific action in momentum space is derived in Sec.~\ref{sec:symmetry}. Depending on the vortex-lattice structure and the Landau-level index $N$, the set $\{\mathbf{k}_i\}$ decomposes into symmetry-related subsets (orbits) under this group. These subsets, indicated by different colors in Fig.~\ref{fig:gap-closures}, are such that all points within a subset can be mapped onto one another by symmetry operations, whereas no symmetry relates points from different subsets. Consequently, all points in a given subset share the same value of $\delta E_N(\mathbf{k}_i)$, while $\delta E_N(\mathbf{k}_i)$ generally differs between subsets.

As a result, the single transition line $\mu = E_N$ splits into two or more distinct lines $\mu = E_N + \delta E_N(\mathbf{k}_i)$.
As $\omega_c / \Delta$ grows, the split lines approach each other, but never merge. The sum of the Chern number jumps across the split lines near $\mu = E_N$ still equals two, but on each individual line the Chern number change can be both positive and negative with an absolute value that can exceed two. As a result, $\mu = E_N$ line actually corresponds to a gapped rather than gapless phase. It can have different Chern numbers for different $N$  including the $\mathcal{C} = 0$ value. In the latter case, it forms a thin sliver of the topologically trivial phase adiabatically connected to the conventional Abrikosov $s$-wave superconductor in the left lower part of the phase diagram.

Splitting of the zeros of $\Delta_{NN}({\bm k})$ into several groups is exemplified for the square and triangular vortex lattices in Fig.~\ref{fig:gap-closures}, which illustrates the result of full calculation performed below in Sec.~\ref{subsec:full-diagonalization}. The points in each panel of Fig.~\ref{fig:gap-closures} are split into groups, which are shown in different colors matching the colors of the transition lines of Figs.~\ref{fig:tr-phase-diagram} and \ref{fig:sq-phase-diagram}. 
Inside each group, the pattern of zeros is the result of the vortex lattice symmetry of the system, see Section~\ref{sec:symmetry}. We list the groups of symmetry-related gap-closure points in Tbl.~\ref{tbl:zeros}. The sum of the topological charges of the zeros in each group equals the Chern number jump across the corresponding transition line, see Sec.~\ref{subsec:jump-matching}. 
\begin{table}
\caption{
The gap closures in the vicinity of the Landau-level lines ($\mu = E_N$) and their degeneracies.}
	\label{tbl:zeros}
	\begin{ruledtabular}
	\begin{tabular}{ccc} 
	& Triangular Lattice & Square lattice \\
	\hline \hline
	$N = 0$ & 2 $\times$ Dirac & 2 $\times$ Dirac \\
    \hline
	$N = 1$ & 2 $\times$ Dirac & 2 $\times$ Dirac \\
	        & 4 $\times$ Dirac & 2 $\times$ QBC \\
    \hline
	$N = 2$ & 4 $\times$ Dirac & 2 $\times$ Dirac \\
	        & 2 $\times$ CBC & 8 $\times$ Dirac  \\
	        & &  8 $\times$ Dirac \\
    \hline
	$N = 3$ & 2 $\times$ Dirac & 2 $\times$ Dirac \\
	        & 12 $\times$ Dirac & 8 $\times$ Dirac \\
	        & 12 $\times$ Dirac  & 2 $\times$ QBC \\
    \hline
	$N = 4$ & 2 $\times$ Dirac & 2 $\times$ Dirac \\
	        & 4 $\times$ Dirac &  8 $\times$ Dirac \\
	        & 12 $\times$ Dirac &  8 $\times$ Dirac \\
	        & 12 $\times$ Dirac & \\
	\end{tabular}
	\end{ruledtabular}
    
    \vspace{0.2cm}
	Dirac: Dirac point, QBC: quadratic band crossing, \\ CBC: cubic band crossing 
\end{table}

The second order perturbation theory is useful to make an estimate of how far the real transition lines diverge from the Landau fan lines $\mu = E_N$ at $\omega_c \gg \Delta$. Taking $N,M \sim 1$, we obtain that 
\be
    \frac{\delta E_N}{\mu} \sim \frac{\Delta^2}{\omega_c \mu} \sim \left( \frac{\Delta}{\omega_c} \right)^2 \qquad (\Delta \ll \omega_c),
\ee
where we used that $\mu \sim \omega_c$ for the lowest Landau levels. 

We emphasize that the zero-Chern phase emerges, surprisingly, even in the limit of a very small $\Delta$ when the chemical potential is tuned to the $N=0,1,2$ Landau level in the case of the triangular vortex lattice, or $N=0,1,3,4$ Landau level in the case of the square vortex lattice. It is remarkable that the topological contribution from quasiparticles near the Fermi level exactly compensates for the filled integer quantum Hall bands below. Even more striking is that for other Landau levels (such as $N=3,4$ for triangular vortex lattice), the superconducting vortex lattice in the middle of a Landau level has a finite Chern number, leading to topological superconductivity no matter how small $\Delta / \omega_c$ is.

\begin{figure}\includegraphics[width=0.49\textwidth]{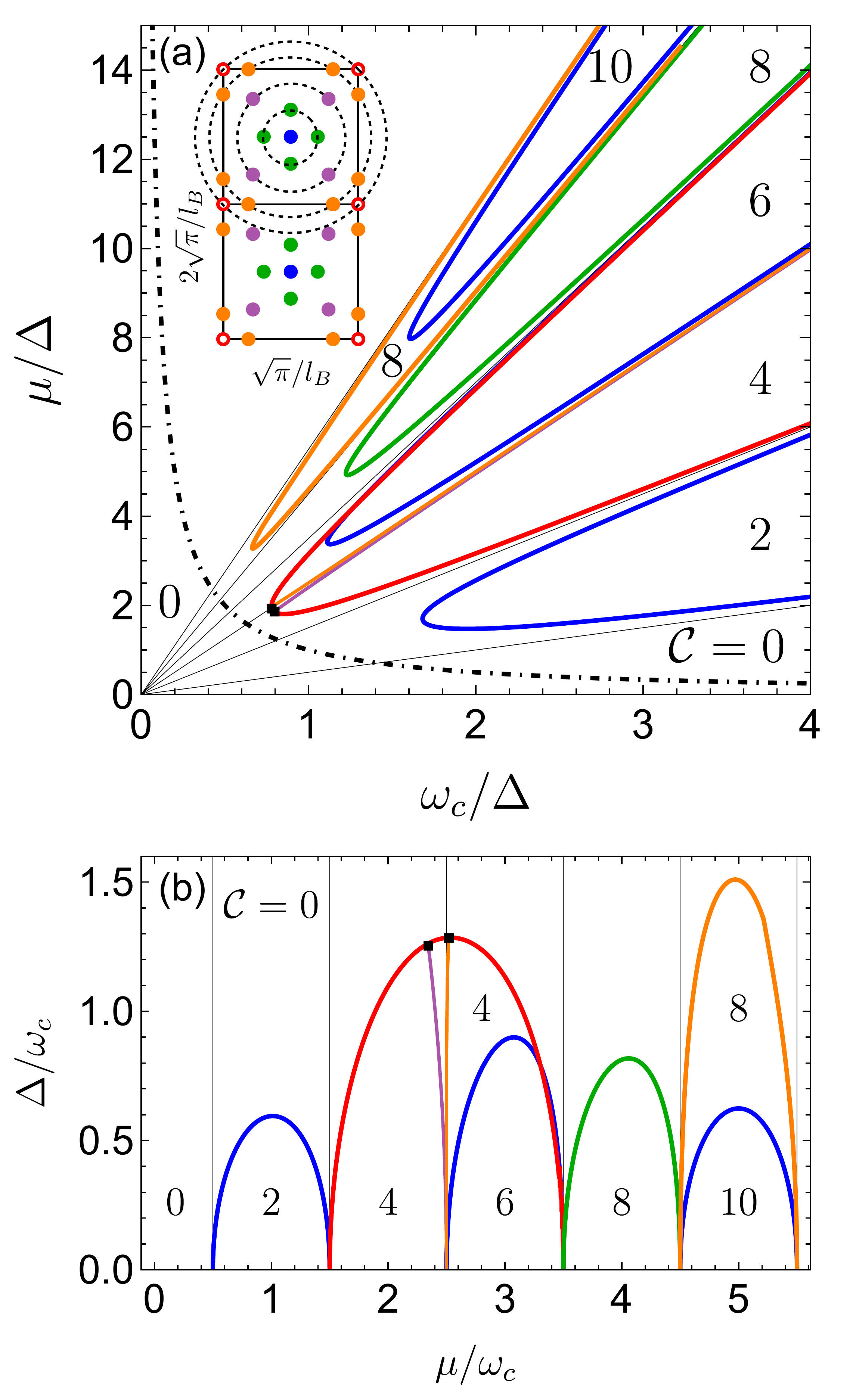}
    \caption{
    \label{fig:sq-phase-diagram}
    (Square vortex lattice) Topological phase diagram of the two-dimensional electron gas in the presence of magnetic field and superconducting order parameter with Abrikosov vortex lattice in two different axes choice. Here $\omega_c$ is the cyclotron frequency, $\mu$ is the chemical potential and $\Delta$ is the superconducting order parameter amplitude. The dashed line ($\omega_c \mu \sim \Delta^2$) is an estimate for the boundary of the region containing topological domes, see Eq.~\eqref{top-trivial-boundary}. The black squares mark the tricritical points.
    Inset: configuration of the degenerate gap closure points in the magnetic Brillouin zone; filled circles mark Dirac points while empty circles stand for the quadratic band touchings. The colors correspond to the colors of the lines in the main plot. 
    As shown in Sec.\ref{sec:symmetry}, blue points coincide with the $C_4$ rotation axes; the other groups of same-color points are symmetric under this operation. When moving along the transition lines in the main plot, the green, orange, and purple points move retaining the $C_4$ symmetry.
    \label{fig:sq-phase-diagram}
}
\end{figure}

\subsection{Arbitrary pairing strength $\Delta / \omega_c$}

\label{subsec:full-diagonalization}
In this Section, we perform a full many-Landau-level calculation.
To compute the quasiparticle dispersion of the BdG Hamiltonian, we substitute the matrix elements \eqref{Delta_NM_k_sq}--\eqref{Delta_NM_k_tr} into \eqref{BdG} obtaining a $\bk$-dependent matrix in the Landau-level index space. We then impose a cutoff $N_{\text{max}} = 30$ (which is sufficient for convergence). 
and diagonalize this matrix numerically for each $\bk$, obtaining the quasiparticle dispersion in the magnetic Brillouin zone. 

We study the obtained quasiparticle spectrum and Bloch wavefunctions using two approaches. The first one is the calculation of the superconducting Chern number given the parameter values ($\mu/\Delta$ and $\omega_c/\Delta$). The expression for the superconducting Chern number is provided by the direct generalization of the (Thouless-Kohmoto-Nightingale-den Nijs) TKNN formula \cite{TKNN, Jain2022helical}:
\be \label{TKNN}
    \mathcal{C} = -\frac{i}{2\pi} \sum_i \int_{\text{BZ}} d^2\bk\, \mathcal{F}_{ii}(\bk),
\ee
where the sum is taken over the occupied bands and $\mathcal{F}_{ii}(\bk)$ is the Berry curvature of the $i$th band:
\be \label{F-BerryCurvature}
    \mathcal{F}_{ii}(\bk) = \epsilon_{lm} \partial_{k_l} \left\langle u_{i,\bk} | \partial_{k_m} | u_{j,\bk}\right\rangle,
\ee
where $\left.|u_{i,\bk}\right\rangle $ is the Bloch wavevector of the $i$th band. Further details on the analytical steps in evaluating \eqref{F-BerryCurvature} for magnetic Bloch wavefunctions, practical implementation of the Brillouin-zone integral in \eqref{TKNN}, and resolving the issue with the hole bands being unbound in energy from below can be found in the Appendix D of Ref.~\onlinecite{MakingSwaveTopological}. 
This method can be used to compute the Chern number at arbitrary points in the $(\omega_c/\Delta,\mu/\Delta)$ plane, and from these data one can construct a finite-resolution phase diagram by evaluating the Chern number on a mesh of points. However, this procedure yields only a coarse image and is not well suited for accurately resolving the shape of the transition lines.

To obtain a higher-resolution phase diagram and determine the transition trajectories more precisely, we employ our second approach to studying topology---an algorithm based on a detailed analysis of gap closures, implemented as follows. First, we study the phase diagram cuts along the slanted lines with fixed ratio of $\mu/\omega_c = N$ (midgap between the Landau levels at small $\Delta$). The gap typically closes several times on this line, see Ref.~\onlinecite{MakingSwaveTopological} for the detailed examples. For each gap closure point on the slanted line, we extend it to the whole dome-shaped topological transition line by iteratively tweaking $\mu/\omega_c$ across the interval $(N - 1/2, N +1/2)$ and adjusting the values of $\Delta$ to maintain the zero quasiparticle gap. That is always possible due to the topological reason: each line delineates two regions with a fixed but different superconducting Chern numbers barring tricritical points. As a result, we obtain a number of topological transition lines, dividing the $\omega_c/\Delta$--$\mu/\Delta$ plane into the regions with different Chern numbers. We already know that it equals zero in the left lower corner and IQHE value, $2 \lfloor \mu / \omega_c + 1/2 \rfloor$, in the outer regions, at small $\Delta$. For the interior domains, we can use the formula \eqref{TKNN} to determine the topology. 

Using the approaches outlined above, we revealed the general behavior of the phase diagram, seen in Figs.~\ref{fig:tr-phase-diagram} and \ref{fig:sq-phase-diagram}: the dome structure of the topologically nontrivial phases, where the domes can both be nested and separated by a sliver of the topologically-trivial phase. In accordance with the results of Ref.~\onlinecite{MakingSwaveTopological}, we observe that at large $\Delta$, corresponding to the left lower corner of Figs.~\ref{fig:tr-phase-diagram}(a) and \ref{fig:sq-phase-diagram}(a), the Chern number is zero, matching the properties of a topologically trivial $s$-wave superconductor with Abrikosov lattice of well-separated vortices. This topologically trivial region is adiabatically connected to the whole $\mu / \omega_c < 1/2$ sector, which extends to the region of large $\omega_c / \Delta$ with no Landau level filled. To estimate the border of the $\mathcal{C} = 0$ corner, we use the following consideration. In the lattice of well-separated vortices, each of them carries Caroli--de Gennes--Matricon bound states, which  have, in the $s$-wave case, a non-zero energy gap. The topology can change once the vortex states start to overlap, i.e. when the wavefunction spread $\xi_v$ becomes of the order of the vortex separation given by $l_B$. After some algebra, that gives the following equation for ther boundary of the topologically trivial region:
\be \label{top-trivial-boundary}
    \mu^* \sim \Delta^2 / \omega_c.
\ee
We show this relation by a dot-dashed in the phase diagrams of Figs.~\ref{fig:tr-phase-diagram}(a) and \ref{fig:sq-phase-diagram}(a). As one can see, this line indeed provides a good estimate for the topologically trivial part of the phase diagram. 
 
\section{Calculating the Chern number through gap closures}
\label{subsec:jump-matching}
Finally, we check that the number and topological charge of the gap closure points ($\mathcal{C}_{\bm{k}^*,\mu^*,\Delta^*}$) on each transition line matches the difference of the Chern numbers in the domains, adjacent to that line ($\mathcal{C}_{1,2}$):
\be \label{Chern-jump-through-points}
    \mathcal{C}_{1} - \mathcal{C}_{2} = \sum_{\bm{k}^*} \mathcal{C}_{\bm{k}^*,\mu^*,\Delta^*}
\ee
To this end, for a chosen parameter-value point on each transition line we determine all gap closure points using a simple minimization algorithm with random initialization. Note that if we move along that transition line, the gap closure points can remain in the same place at a high-symmetry point, move along a high-symmetry line, or, for a general point, move together with its symmetry-related counterparts retaining their $C_{4(6)}$ symmetry with respect to the closest rotation center. As explained above, we observe both Dirac and higher-order gap closure points. To reliably determine their topological charge, we surround each gap closure point $(k_x^*, k_y^*)$ point at parameters $\mu^*$ and $\Delta^*$ (measured in the units of $\omega_c$) with a small cube in the $(k_x, k_y, \Delta)$ space and evaluate the Berry curvature flux of one of the touching bands through this cube:
\be
    \mathcal{C}_{\bm{k}^*,\mu^*,\Delta^*} = \int_{\text{\mancube}} F dS,
\ee
where $F$ is the 2-form notation for the Berry curvature \eqref{F-BerryCurvature}, where $\mu$ takes the role of the third momentum component. The flux through each face of the cube is evaluated by tiling it with small square plaquettes and applying the standard formula from Ref.~\onlinecite{Fukui-Berry}. Note that the absolute value of the topological charge determines the power-law of the quasiparticle dispersion relation in the vicinity of this point: $E\propto \delta k^{|\mathcal{C}|}$.

We found Dirac and quadratic band crossing points in the square vortex lattice (see caption to Fig.~\ref{fig:sq-phase-diagram} for details) and Dirac and cubic band crossing points for the triangular lattice (see caption to Fig.~\ref{fig:tr-phase-diagram}). In general, we observed agreement between the Chern number values in each domain evaluated through \eqref{TKNN} and the sum of the topological indices of each point at that closure in accord with Eq.~\eqref{Chern-jump-through-points}. Moreover, in regions of the phase diagram where transition lines come close to each other and for small patches of a phase with a distinct Chern number, the Brillouin-zone integration of the Berry curvature becomes numerically unreliable. In these cases we rely on the second approach: we determine the Chern number in a given domain by starting from the Chern number of an adjacent domain and adding the sum of the topological charges of all gap closings on the transition line.
Overall, that allowed us to reliably establish the rich phase diagram of the topology of the quantum Hall-superconductor transition in the mean-field approximation. 

\section{Space group of the BdG quasiparticles in the superconducting vortex lattice}
\label{sec:symmetry}

In this Section, we explain the symmetries of the quasiparticle dispersion in the magnetic Brillouin zone and how they protect the degeneracy of the gap closures. These symmetry considerations are applicable both to the full spectrum of the BdG Hamiltonian \eqref{BdG} that we study in Section \ref{sec:many-LLs} and the single-Landau level limit, discussed in the Section \ref{sec:single-LL}. 

The symmetries that we study include both magnetic translations as well as rotations and mirrors forming the point group of the superconducting vortex lattice. Importantly,  the presence of $h/2e$ flux quanta per unit cell leads to non-commutativity between magnetic lattice translations and requires a careful treatment different from the standard space group theory~\cite{Cosic-Hofstadter-groups, JCano-flux}. 
Symmetry analysis in our setup is closely related to the Hofstadter phenomenon\cite{Harper, Hofstadter} with $q=2$ since a vortex lattice unit cell contains a superconducting flux quantum $h/(2e)$, i.e., one half of the electron/hole magnetic flux quantum~\cite{Shaffer-HofstadterSC}. 

Unlike previous works, our analysis is tailored to a continuum Landau-level BdG problem with an imposed Abrikosov vortex lattice, and uses the vortex-lattice symmetries combined with the BdG-specific half-cell translation structure to explain the multiplicities of gap closings and Chern jump sizes. We also take into account particle-hole symmetry inherent in BdG formalism and combination of lattice and time-reversal symmetry, both of which play important roles below. 

\subsection{Particle-hole symmetry}

BdG Hamiltonian \eqref{BdG} by construction, satisfies the particle-hole symmetry:
\be 
    \mathcal{H}_{\text{BdG}}(\bm{k}) = - \tau_x \mathcal{H}_{\text{BdG}}^*(-\bm{k}) \tau_x,
\ee
where $\tau_x$ is the Pauli matrix in the Nambu (particle-hole) space. In the momentum space, particle-hole symmetry acts by replacing the quasimomentum by the opposite value, $\bm{k} \rightarrow - \bm{k}$, and flipping the sign of the quasiparticle energy, $E\rightarrow-E$. Therefore, each zero at $\bm{k} \neq 0$ is accompanied by a zero at the opposite quasimomentum. Therefore, the patterns of zeros satisfies the inversion symmetry with respect to the $\bk=(0,0)$ point, corresponding to the lower left corner of the Brillouin zone, as depicted in Fig.~\ref{fig:Lattices}(c, d). By combining this operation with the translation by the reciprocal lattice vectors $\bG_1$, $\bG_2$, and their sum, one concludes that the center of the the Brillouin zone (as well as all its edge centers) are also inversion centers as can be seen in Fig.~\ref{fig:gap-closures}.

\subsection{Translation by half of the magnetic unit cell}
\label{sec:half-magnetic-translation}

In this Section, we describe the symmetry that guarantees that each gap closure has a pair in the other half of the Brillouin zone; hence, the superconducting Chern number always changes by an even integer. 

As mentioned earlier, wavefunctions \eqref{wavefunctions_magnetic_BZ} have a period $2a_x$ in the $x$-direction in contrast with the original vortex lattice that has a period $a_x$. In other words, translation by the vector $(a_x, 0)$ maps the Hamiltonian onto itself, while the wavefunction $\psi^{e(h)}(\bm{k})$, see Eq.~\eqref{wavefunctions_magnetic_BZ}, transforms into the wavefunction at a shifted quasimomentum:
\be \label{half-magnetic-transform}
    \psi^{e(h)}_{\bm{k}}(x + a_x, y) = e^{i k_x a_x} \psi^{e(h)}_{\bm{k} + \bG_2 / 2} (x,y),
\ee
where $\bG_2 / 2$ is a half of the $y$-direction reciprocal vector, both for square and triangular vortex lattice. Thus, each of the Brillouin zone in Fig.~\ref{fig:Lattices} can divided into two equal parts with identical dispersion. Therefore, each gap closure point (Dirac or higher-order) has its counterpart in the opposite half of the Brillouin zone. Correspondingly, topological Chern number across the transition changes always changes by an even number, as was discussed in Ref.~\onlinecite{MakingSwaveTopological} and confirmed by the results of Section~\ref{sec:many-LLs}.
To sum up, Eq.~\eqref{half-magnetic-transform} proves that having one half of the magnetic flux in a unit cell ensures twofold degeneracy of all states in the energy spectrum.

\subsection{Rotation symmetry}
Degeneracies of the gap closure points are further enhanced by the lattice symmetries of the superconducting vortex lattice correspondingly. They include $C_4$ and $C_6$ rotation symmetries of the square and triangular lattice. Below, we derive how these symmetries manifest at the level of the quasiparticle dispersion and use them to explain the patterns of zeros in Fig.~\ref{fig:gap-closures} and insets in Fig.~\ref{fig:sq-phase-diagram} and \ref{fig:tr-phase-diagram}. 

The amplitude $|\Delta(r)|$ of the superconducting vortex lattice is symmetric with respect to the rotation symmetry, $\R$, where $\R = C_4$ for the square vortex lattice and $\R = C_6$ for the triangular lattice. However, such rotation does not preserve the phase of $\Delta(r)$ as well as the Landau gauge for the magnetic vector potential  $\bA(r) = (By,0,0)$. To return the Hamiltonian to the original representation, one has to apply a gauge transform $\Delta \rightarrow \Delta e^{2i\chi(\br)}$, $\bA \rightarrow \bA + \nabla \chi(\br)$. 

The expression for $\chi(\br)$ in the square-lattice case and $\pi/2$-rotation is
\be
    \chi^{\text{sq}}(\br) = \frac{xy}{l_B^2},
\ee
while for the triangular lattice case rotated by $\pi/3$,
\be
    \chi^{\text{tr}}(\br) = \frac{\sqrt{3}}{8l_B^2}(x^2 - y^2) + \frac{3xy}{4l_B^2}.
\ee
Applying the gauge transform with the function $\chi(\br')$ brings $\Delta$ and $\bA$ to their original form:
\be
    \Delta(\R^{-1} \br') e^{2i \chi(\br')} = \Delta(\br'),
\ee
\be
    \R \bA(\R^{-1}\br') - \nabla \chi(\br') = \bA (\br') .
\ee
Under the same composite operation, the wavefunctions transform into the expression
\be \label{psi-rotation-transform}
    \tilde{\psi}_{\bm{\bm{k}}}^{e(h)}(\bm{r}') = \psi_{\bm{k}}^{e(h)}(\R^{-1} \br') e^{\pm i \chi(\bm{r'})}.
\ee
By the symmetry, the transformed function $\tilde{\psi}_{N, \bk}^{e(h)}$ is also a solution of the BdG Hamiltonian bearing the same energy $\epsilon_{\bm{k}}$. However, direct algebraic manipulation with Eq.~(\ref{psi-rotation-transform}) shows that $\tilde{\psi}_{N, \bk}^{e(h)}$ generally satisfies a different magnetic boundary conditions compared to those for $\psi_{\bm{\bm{k}}}^{e(h)}(\bm{r})$, see Eqs.~\eqref{psi_translations_sq} and \eqref{psi_translations_triangular}. This is expected, since the magnetic BZ has the rectangular (square-lattice) or parallelogram (triangular-lattice) shape, which is not invariant under the rotation symmetry. 
To construct the rotated wavefunctions that satisfy the same boundary conditions, one has to use the symmetry \eqref{half-magnetic-transform}, representing translation by a half of the magnetic unit cell. It guarantees that $\psi_{N, \bk}^{e(h)}$ and $\psi_{N, \bk + \bG /2}^{e(h)}$ are both eigenstates of the Bloch Hamiltonian with the same energy; the same is valid for their transforms $\tilde{\psi}_{N, \bk}^{e(h)}$ and $\tilde{\psi}_{N, \bk + \bG /2}^{e(h)}$ under the transformation \eqref{psi-rotation-transform}. We now consider the linear combinations $\tilde{\psi}_{\bm{\bm{k}}}^{e(h)} \pm  \tilde{\psi}_{\bm{\bm{k}+\bG_2}}^{e(h)}$. After some algebra, one can show that they  satisfy the original boundary conditions \eqref{psi_translations_sq} and \eqref{psi_translations_triangular} with a certain quasimomentum. Explicitly,
\begin{subequations}
\label{rotated-wavefunctions}
\begin{align}
    \label{rotated-wavefunctions-1}
    \tilde{\psi}_{\bm{\bm{k}}}^{e(h)} + \tilde{\psi}_{\bm{\bm{k}+\bG_2}}^{e(h)} & = \psi_{N, \R\bk + \beta \bG_2 / 2}^{e(h)}\\
    \tilde{\psi}_{\bm{\bm{k}}}^{e(h)} - \tilde{\psi}_{\bm{\bm{k}+\bG_2}}^{e(h)} & =\psi_{N, \R\bk + \beta \bG_2 / 2 + \R \bG_2/2}^{e(h)} .
\end{align}
\end{subequations}
In particular, Eq.~\eqref{rotated-wavefunctions-1} means that the quasiparticle dispersion satisfies the condition $E_{\bk} = E_{\tilde{\bk}}$ with
\be \label{rotated-k}
    \tilde{\bk} = \R\bk + \beta \bG_2 / 2,
\ee
where $\beta = 0$ for the square vortex lattice and $\beta = 1$ for the triangular lattice. As a result, quasiparticle dispersion and the pattern of zeros of the gap closure points satisfy the rotation symmetry $C_{4(6)}$ for the square(triangular) lattice cases with respect to the certain rotation points. Combining \eqref{rotated-k} with the reciprocal lattice translations, one can show that the center of each half of the Brillouin zone in Figs.~\ref{fig:Lattices}(c,d) is a rotation center (shown with stars).

\subsection{Space-time reflection symmetry}

The superconducting order parameter is invariant under the vertical mirror symmetry $\mathcal{M}$ of the vortex lattice combined with the time-reversal symmetry $\mathcal{T}$; no additional gauge transform is needed. 
The action of this symmetry on the quasiparticle dispersion and the pattern of the gap closure points is straightforward---they are symmetric with respect to the constituent mirror acting in the momentum space. The reflection line can be chosen (by combining it with the reciprocal vector translations) to pass through the magnetic Brillouin zone center, as shown with the dashed line  in Fig.~\ref{fig:Lattices}(c, d). Combining this mirror with rotations generates the rest of the mirror operations with the symmetry centers indicated by the stars in Fig.~\ref{fig:Lattices}(c, d). Mirror symmetries are important to explain some of the $8$- and $12$-fold degeneracies, which we do in the following Section.

\subsection{Patterns of the gap closure points}
\label{secsec:patterns}

Having established the symmetries of the quasiparticle dispersion, we now analyze the patterns of the gap closure points. 
The latter are shown in Fig.~\ref{fig:gap-closures} for the single-Landau level limit. As explained above, upon inclusion of the coupling to the other Landau levels in the full BdG calculation, the gap closure points break down into groups shown in different colors in the insets of Figs.~\ref{fig:tr-phase-diagram} and \ref{fig:sq-phase-diagram}. Additionally, in Fig.~\ref{fig:another-12-pattern} in Appendix we show a highly-degenerate gap closure pattern in the full calculation for $\mu = E_{14} + \omega_c / 2$.  The points shown in one color occur at the same values of the parameters $\mu / \Delta$ and $\omega_c / \Delta$ and are all related to each other by symmetries.

First, we reiterate that in all considered cases the Brillouin zone can be divided in two equivalent halves as shown in Fig.~\ref{fig:Lattices}(c,d) so that each gap closure point has a counterpart in the other half. 

Second, one can observe that gap can close both at high-symmetry points (in the corners of the half-BZ and the centers of its halves), at points pinned to the high-symmetry lines (orange dots in the inset to Fig.~\ref{fig:sq-phase-diagram}), and at general points. In the first case, the point can be a higher-order band touching rather than a Dirac node and contribute a larger integer to the Chern number jump across the transition. We observed quadratic band touchings for the square-lattice case (red circles in the inset to Fig.~\ref{fig:sq-phase-diagram}) and cubic band touchings in the triangular-lattice case (red crosses in the inset to Fig.~\ref{fig:tr-phase-diagram}).

Next, all points respect the aforementioned rotation and mirror symmetries. Considering a single rotation point [a star in Fig.~\ref{fig:Lattices}(c,d)] and a set of mirror symmetries passing through this point, 
we observe that the orbits of this group action can contain $1$, $4$, and $8$ points in the square-lattice case and $1$, $6$, and $12$ points in the triangular-lattice case, as expected from our general space group analysis for superconducting vortex lattice.

As mentioned above, the Chern number jump at any transition line equals the sum of the topological charges of each gap-closure point (Dirac, QBC, or CBC) in the magnetic Brillouin zone. However, determining the number of such points in the Brillouin zone is nontrivial and one should take care when applying the vortex-lattice symmetries for that purpose.
Importantly, there is \textit{no} simple relation between the number of points in the naive orbit and the  Chern number jump across the corresponding transition. Instead, we observed several distinct cases. The first case is illustrated by the blue, orange, and purple points in Fig.~\ref{fig:tr-phase-diagram} and the blue and green points in Fig.~\ref{fig:sq-phase-diagram}. Here, the number of points in one half of the BZ equals the length of the point group orbit; the total number of points is twice as large and matches the Chern number jump. Similarly, each of the two red crosses in Fig.~\ref{fig:tr-phase-diagram} form an orbit of length one in each BZ half. Since they correspond to a qubic band crossing, the Chern number jump is  $2\times3 = 6$. 

The second case is exemplified by the orange points and empty red circles in Fig.~\ref{fig:sq-phase-diagram}. While the naive orbit lengths are eight and four, respectively, some of these points are equivalent under translation by a reciprocal lattice vector. They should not be double-counted, so the Chern number jump is smaller than the doubled orbit length. Explicit counting of the gap-closure points is presented in App.~\ref{app:counting-gap-closures}.

Finally, the third case corresponds to the green points in Fig.~\ref{fig:tr-phase-diagram}. There are six points in the point group orbit, but only three lie within the selected BZ. Moreover, the orbits for the two symmetry centers intersect, so, as shown in Fig.~\ref{fig:tr-phase-diagram}, the total number of the gap closures in the full BZ is four. This case is especially interesting, as it demonstrates that a Chern number jump of four can occur on the triangular vortex lattice, even though the $C_6$ and $D_6$ point groups have no orbit of that length. Overall, the combination of large symmetry group orbits and their nontrivial intersection with the Brillouin zone give rise to the rich structure seen in the phase diagrams of Figs.~\ref{fig:tr-phase-diagram} and \ref{fig:sq-phase-diagram}, which include topological transition lines with Chern number jumps of $1$, $2$, $4$, $6$, $8$, and $12$. Our analysis also suggests that a jump of $24$ is possible in the triangular lattice case, although we did not observe it in the portion of the phase diagram explored.

\section{Effects of the vortex lattice distortion}
\label{sec:lattice-distortion}

In the current Section, we study the distortion of the vortex lattice lattice that lowers its symmetry. We demonstrate that it leads to the further splittings of the transition lines.
\begin{figure}[h]
	\includegraphics[width=0.48\textwidth]{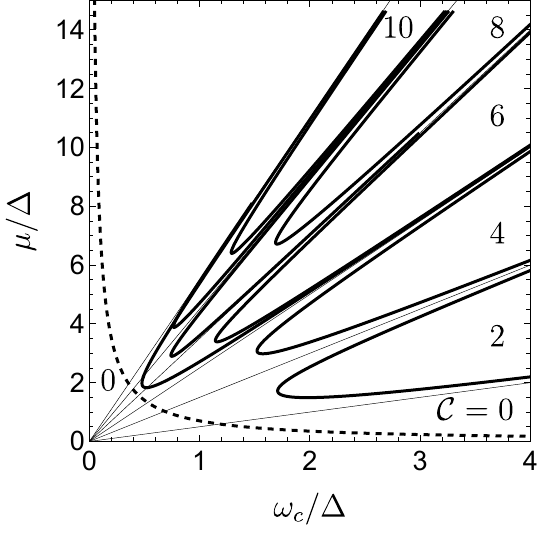}\caption{
    (Anisotropic triangular vortex lattice) Topological phase diagram for the anisotropically stretched triangular vortex lattice, compare with the Fig.~\ref{fig:tr-phase-diagram} for the undistorted case. Reduction of the rotation symmetry from $C_6$ to $C_2$ leads to the splitting of the transition lines with a high Chern number jump into several lines with a smaller one. 
	\label{fig:aniso-tr-phase-diagram}}
\end{figure}

As explained in the previous Sections, topological transitions in the quasiparticle band structure formed by a square or triangular vortex lattice involve gap closure at the transition lines in multiple points (up to 12 points) in the Brillouin zone.
Such high degeneracy is protected by the rotation symmetry of the lattice and relies on its unperturbed structure. In this Section, we show that stretching the triangular lattice in one direction (and compressing it in the orthogonal one) reduces the gap closure degeneracy through splitting of the transition lines into lines with a smaller jump of the Chern number.  

We consider now the deformed triangular lattice, for which $\Delta(r)$ still given by \eqref{Delta_r}, but with $a_x = a_y = \sqrt{\pi} l_B^2$ (remember that the $C_6$-symmetric triangular lattice had $a_x^{\text{tr}} = \sqrt{2\pi/\sqrt{3}}$ and $a_y^{\text{tr}} = \sqrt{\sqrt{3}\pi/2}$, see Section~\ref{sec:model} and Fig.~\ref{fig:Lattices}). We now repeat the calculations of the Section~\ref{sec:many-LLs} with modified values if $a_x$ and $a_y$, evaluating the quasiparticles dispersion, superconducting Chern number and tracing the topological transition lines on the $\{\omega_c/\Delta,\,\mu/\Delta\}$ phase diagram. The results are presented in Fig.~\ref{fig:aniso-tr-phase-diagram}. 
An immediate observation is that there are no longer transitions where the Chern number jumps by $6$ or $12$. It aligns with the predictions from the group theory analysis: since the stretched vortex lattice lacks the $C_6$ rotation symmetry (which is reduced to the $C_2$ symmetry), there are no orbits of the group action that contain $6$ (or multiples of $6$) points. Consequently, the transition lines that correspond to such Chern number jumps are split and we observe more transitions in Fig.~\ref{fig:aniso-tr-phase-diagram} compared to Fig.~\ref{fig:tr-phase-diagram}. At the same time, the system still demonstrates Landau-fan behavior in the $\mu \omega_c / \Delta^2 \gg 1$ limit, while in the opposite limit $\mu \omega_c / \Delta^2 \ll 1$ topologically trivial ($\mathcal{C} = 0$) Abrikosov $s$-wave superconducting state is rendered. 

\section{Discussion}

In the present work, we obtained a comprehensive
topological phase diagram of a two-dimensional spin-degenerate electron gas in the presence of superconducting vortex-lattice correlations accounting for the intermixing of many Landau levels. The most prominent result is that the Landau-fan transition lines of the Integer Quantum Hall Effect, $\mu = E_N$, split into several distinct transition lines (with details depending on $N$) that approach each other in the limit $\omega_c \gg \Delta$ but never merge, giving rise to intermediate superconducting phases with distinct topology, see Figs.~\ref{fig:tr-phase-diagram} and \ref{fig:sq-phase-diagram}. Instead of a single IQHE Chern-number jump of one per spin projection, the Chern number now jumps multiple times, with individual jumps that can be relatively large even integers (up to 12 observed), and of either sign. In the regime $\Delta / \omega_c \sim 1$, dome-shaped topological domains appear. A further increase of $\Delta / \omega_c$ drives the system into the topologically trivial Abrikosov $s$-wave superconducting phase as  shown in Ref.~\onlinecite{MakingSwaveTopological}.

Remarkably, our results show that in the opposite limit of $\Delta/\omega_c \ll 1$ and chemical potential tuned to a Landau level energy (see Sec.~\ref{sec:many-LLs} for the details), the Chern number may be zero, making the state adiabatically connected to the large-gap, $s$-wave superconductor limit, or may be finite.
In particular, proximitizing any of the lowest three Landau levels, $N=0$--$2$, with a triangular superconducting vortex lattice leads to a topologically trivial phase, $\mathcal{C} = 0$, while for the $N = 3$ Landau level, a topologically nontrivial phase, $\mathcal{C} = 6$, emerges---even in the limit of arbitrarily small $\Delta/\omega_c$. We conclude that, in experiments on proximitized two-dimensional electron systems, it is worth going beyond the lowest Landau levels in the search for topological superconducting phases.

The large Chern-number jumps at the split transition lines occur due to multiple gap closures in the magnetic Brillouin zone, which can be Dirac or higher-order band crossings. This degeneracy is protected by the crystal symmetries of the vortex lattice, cf. Figs.~\ref{fig:Lattices} and \ref{fig:gap-closures}, including its point group. It is therefore expected that perturbations breaking the vortex-lattice symmetry will produce additional splitting of the transition lines. We tested this expectation by considering a distortion of the triangular vortex lattice in Sec.~\ref{sec:lattice-distortion}, which indeed led to further splitting of the topological transitions.

Disorder is another mechanism that reduces perfect lattice symmetry. 
We therefore expect potential disorder to produce maximal splitting of the transition lines in the phase diagram for each fixed disorder realization, such that every line carries a Chern jump of two, which is still protected by the spin degeneracy. 
Disordered quantum Hall systems are also known to showcase floating (levitation) of the levels \cite{Khmelnitsky, LaughlinLevitation}. We expect that the topological transitions revealed in the current work levitate together with them being ``pinched'' between the $\Delta = 0$ topological transition lines. 
A detailed study of the disordered system represents a separate interesting problem; see also Ref.~\onlinecite{Burmi-new}.

In our calculation, we neglected the Zeeman coupling of electrons to the magnetic field. Including it will modify the phase diagram but will preserve the evenness of the Chern-number jumps due to the spin-rotation symmetry. By contrast, including both Zeeman splitting and spin-orbit coupling brings the system to class D\cite{AltlandZirnbauer} and allows for odd Chern-number jumps and Majorana chiral modes. This scenario has been studied previously \cite{Jain2022helical, McDonald_new, MishmashYazdaniZaletel}, though mostly in the limit of one or only a few Landau levels.

\section{Acknowledgments}
This work was supported at Yale University by NSF Grant No. DMR-2410182 and by the Air Force Office of Scientific Research (AFOSR) under Award No. FA95502510287.

\appendix 

\section{Magnetic Bloch states}
\label{app:magnetic-bloch-states}

In this Appendix, we present technical details regarding the superconducting pairing potential $\Delta(r)$ and magnetic Bloch wavefunctions. We start by writing Eq.~\eqref{Delta_r} for $\Delta(r)$ explicitly as \cite{Abrikosov_vortex_lattice}
\be 
	 \Delta(\br) = 2^{1/4} \Delta \sum_t C_t \phi_{0, \sqrt{2} t a_y} (\sqrt{2}\br),
\ee
where $\phi_{N, k_x l_B^2}$ is $N$th Landau level wavefunction in the Landau gauge with the quasimomentum $k_x$ in the $x$-direction and center-of-orbit shift $k_x l_B^2$ in the $y$-direction. The coefficient $\sqrt{2}$ is due to the Cooper pair charge $2e$. As explained in the main part, $C_t = 1$ for the square lattice and $C_t = \exp (-i \pi t^2 / 2)$ for the triangular lattice. The absolute value $|\Delta(r)|$ is a periodic function with the principal vectors $\bm{a}^{\text{sq}} = (a_x^{\text{sq}},0)$, $\bm{b}^{\text{sq}} = (0, a_y^{\text{sq}})$ with $a_x^{\text{sq}} = a_y^{\text{sq}} = \sqrt{\pi}l_B$ for the square lattice and $\bm{a}^{\text{tr}} = (a_x^{\text{tr}},0)$, $\bm{b}^{\text{tr}} = (a_x^{\text{tr}}/2, a_y^{\text{tr}})$ with $a_x^{\text{tr}} = \sqrt{2\pi/\sqrt{3}}$, $a_y^{\text{tr}} = \sqrt{\sqrt{3}\pi/2}$ for the triangular lattice, see Fig.~\ref{fig:Lattices}(a,b).

Further analysis of the BdG Hamiltonian \eqref{BdG} can be performed by going to the basis of magnetic Bloch wavefunctions:
\be \label{wavefunctions_magnetic_BZ}
	\psi_{N, k_x, k_y}^e (\br) = \sqrt{\frac{a_{y}}{L_{y}}} \sum_t \sqrt{C_t} e^{i k_y a_y} \phi_{N, k_x l_B^2 + a_y t} (\br),
\ee
where
\be \label{phi-through-varphi}
    \phi_{N, k_x l_B^2 + a_y t} (\br) = e^{ik_x x + i \pi t x / a_x} \varphi_{N} (y - k_x l_B^2 - a_y t). 
\ee
From Eq.~\eqref{phi-through-varphi}, it is evident that the wavefunction \eqref{wavefunctions_magnetic_BZ} has period $2a_x$ rather than $a_x$ in the $x$-direction. In other words, the magnetic unit cell [highlighted in gray in Fig.~\ref{fig:Lattices}(a,b)] is twice as large as the vortex-lattice unit cell. Such doubling of the Bogoliubov quasiparticle unit cell with respect to the unit cell of the vortex lattice is well-known \cite{Franz-Tesanovic, Liu-Franz} and 
follows from the fact that a Cooper pair carries charge-$2e$ and correspondingly each vortex hosts a superconducting flux quantum $h/2e$.  
Consequently, the magnetic Brillouin zone of quasiparticles is two times smaller, see Fig.~\ref{fig:Lattices}(c,d). Explicitly, the reciprocal unit vectors $\bG_1$ and $\bG_2$ are given by $\bG_1^{\text{sq}} = \left(\sqrt{\pi} / l_B,0 \right)$ and $\bG_2^{\text{sq}} = \left(0,2 \sqrt{\pi} / l_B \right)$ for the square lattice and $\bG_1^{\text{tr}} = \left(\sqrt{\sqrt{3}\pi/2}/l_B, -\sqrt{\pi/(2\sqrt{3})} /l_B \right)$ and $\bG_2^{\text{tr}} = \left(0, 2\sqrt{2\pi/\sqrt{3}}/l_B\right)$. 

The wavefunctions \eqref{wavefunctions_magnetic_BZ} satisfy generalized (quasiperiodic) boundary conditions in the Brillouin zone in accordance with the action of the magnetic translation group. The explicit expressions for the boundary conditions are:
\begin{subequations}
\label{psi_translations_sq}
\begin{align}
    & \psi_{N, \bk}^{e(h)} (x + 2 a_x, y) = e^{2 i k_x a_x} \psi_{N, \bk}^{e(h)} (x, a_y), \\ 
    & \psi_{N, \bk}^{e(h)} (x, y + a_y) = e^{\pm i \pi x / a_x} e^{i k_y a_y} \psi_{N, \bk}^{e(h)} (x, a_y)
\end{align}
\end{subequations}
in the case of a square lattice, and
\begin{subequations}
\label{psi_translations_triangular}
\begin{align}
    & \psi_{N, \bk}^{e(h)} (x + 2 a_x, y) = e^{2 i k_x a_x} \psi_{N, \bk}^{e(h)} (x, a_y), \\ 
    & \psi_{N, \bk}^{e(h)} (x + a_x / 2, y + a_y) = e^{i k_x a_x / 2 + i k_y a_y \pm i \pi /4 \pm i \pi x / a_x} \nonumber \\
    &  \qquad \qquad \qquad \qquad \qquad  \qquad \qquad \ \ \times\psi_{N, \bk}^{e(h)} (x, a_y)
\end{align}
\end{subequations}
for a triangular vortex lattice.

The next stage of our calculation is to evaluate the matrix elements of $\Delta(r)$ between the wavefunctions \eqref{wavefunctions_magnetic_BZ}. The result \cite{McDonald2} reads:
\be \label{Delta_Landau_basis}
	\left\langle \psi_{N,\bk}^e |\Delta(\br) | \psi_{M,\bk'}^h \right\rangle = \delta_{\bk, \bk'} \Delta_{NM} (\bk).
\ee
For the square lattice, the coefficients $\Delta_{NM}(k)$ are:
\be \label{Delta_NM_k_sq}
    \Delta_{NM}^{\text{sq}}(\bk) = \Delta  \sum_{t = -\infty}^{\infty} c_0 e^{2 i k_y a_y t} \varphi_{M + N}(2t a_y + 2k_x l^2),
\ee
where $c_0 = (-1)^N 2^{-(M+N)/2} \sqrt{C_{M+N}^M}$ and $\varphi_{N}(y)=\sqrt{1/(2^{N}N!\sqrt{\pi}l_B)} H_{N}(y / l_B)e^{-y^{2}/(2l_B^{2})}$, with $H_N$ being the Hermite polynomials. For the triangular lattice, we find
\be \label{Delta_NM_k_tr}
    \Delta_{NM}^{\text{tr}}(\bk) = c_0 \sum_{t = -\infty}^{\infty} e^{2 i k_y a_y t - i \pi t^2 / 2} \varphi_{M + N}(2t a_y + 2k_x l^2).
\ee
Our many-Landau-level analysis is based on diagonalizing the Hamiltonian \eqref{BdG} with the matrix elements Eq.\eqref{Delta_NM_k_sq} or Eq.~\eqref{Delta_NM_k_tr} and studying the Chern numbers of the resulting bands as well as the nature of the topological transitions between them.

\section{
The $\Delta \mathcal{C}=12$ topological transitions: $\nu=10$ vs. $\nu=28$}

\begin{figure}[h]
\includegraphics[width=0.2\textwidth]{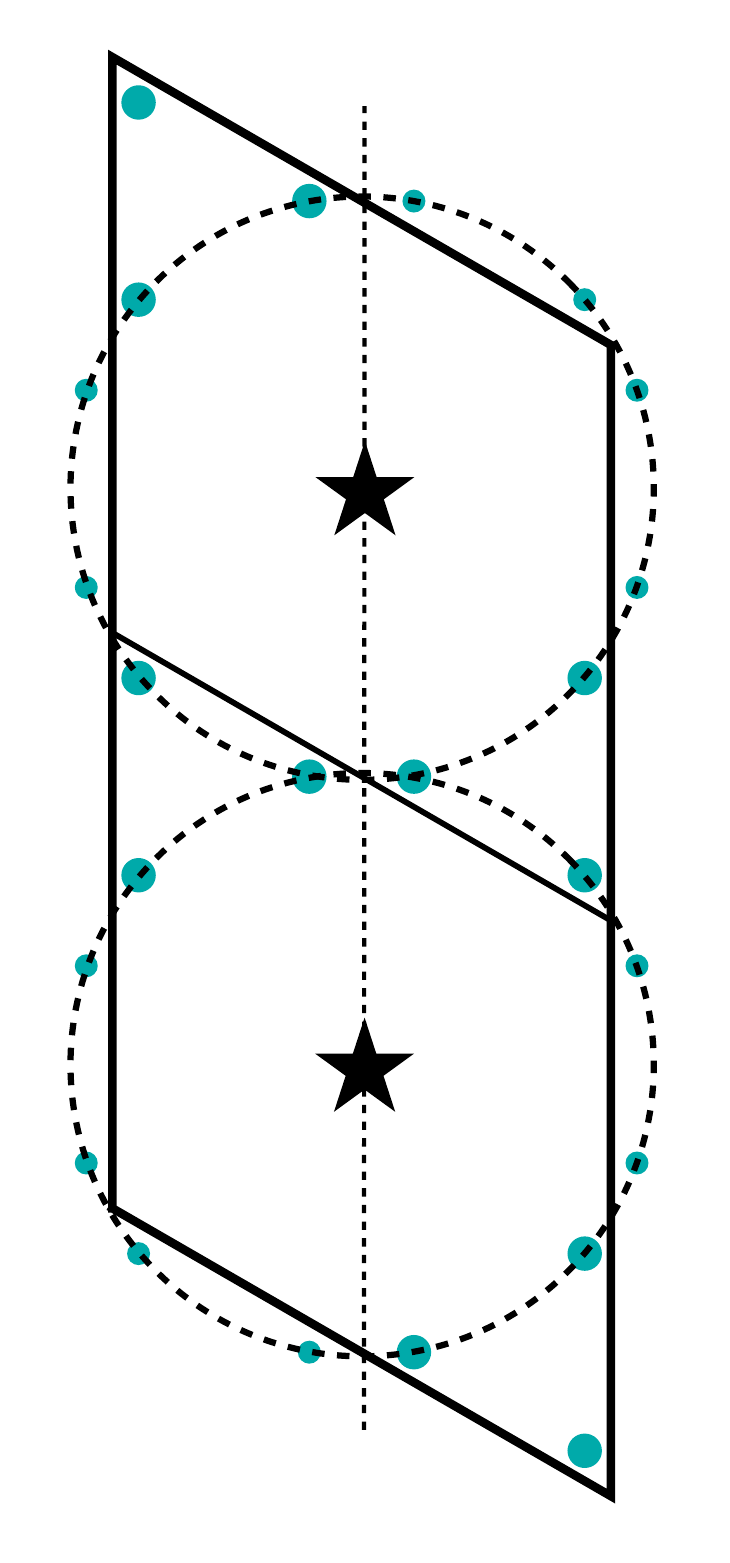}
	\caption{(Triangular vortex lattice) Gap closure points that represent a different, compared to the orange and purple dots in the inset of Fig.~\ref{fig:tr-phase-diagram}, realization of the $\Delta \mathcal{C}=12$ topological transition. Such configuration occurs at the filling factor $\nu=28$, studied in Ref.~\onlinecite{MakingSwaveTopological}. For a single symmetry center, the orbit contains $12$ gap closure points, protected by a combination of rotation and mirror symmetries. 
    \label{fig:another-12-pattern}
    }
\end{figure}
In Sec.~\ref{subsec:full-diagonalization} we studied in detail the phase diagram up to the filling factor of $\nu = 10$. The realization of the topological transitions through multiple gap closure points was illustrated in the insets of Figs.~\ref{fig:tr-phase-diagram} and \ref{fig:sq-phase-diagram}. In particular, for the triangular vortex lattice, we observed a Chern number jump by $12$ at orange and purple lines in Fig.~\ref{fig:tr-phase-diagram}. At these lines, there are six symmetry-related gap closure points in each half of the Brillouin zone, totaling to the $12$ points in full magnetic BZ. 

In this Appendix, we demonstrate another occurrence
of a $\Delta \mathcal{C}=12$ Chern number jump with a pattern of the gap-closure points protected by a combination of rotation and mirror symmetries. The symmetry analaysis we present here explains the $\Delta \mathcal{C}=12$ transition at fixed filling factor $\nu = 28$ which we found earlier in Ref.~\onlinecite{MakingSwaveTopological}.
The pattern of the simultaneous gap closures for this transition is shown in Fig.~\ref{fig:another-12-pattern}. As one can see, there are $12$ symmetry-related points;  protection of this degeneracy requires both mirror and rotation operations. Including rotations and reflections with respect to both symmetry centers (shown with stars in Fig.~\ref{fig:Lattices}(d)) as well as translations in the reciprocal space leads to an intricate pattern with a total of $12$ Dirac points in the magnetic BZ, demonstrated in Fig.~\ref{fig:another-12-pattern}.
\newpage

\begin{widetext}
\section{
Counting of the gap-closure points}
\label{app:counting-gap-closures}

In the Sec.~\ref{secsec:patterns}, we explained that one should use care when counting the number of gap-closure points in the magnetic Brillouin zone. Here we show how exactly this counting works, see Fig.~\ref{fig:gap-closures-numbered} that complements Fig.~\ref{fig:gap-closures} of the main text. 
\begin{figure}[h]
\includegraphics[width=0.75\textwidth]{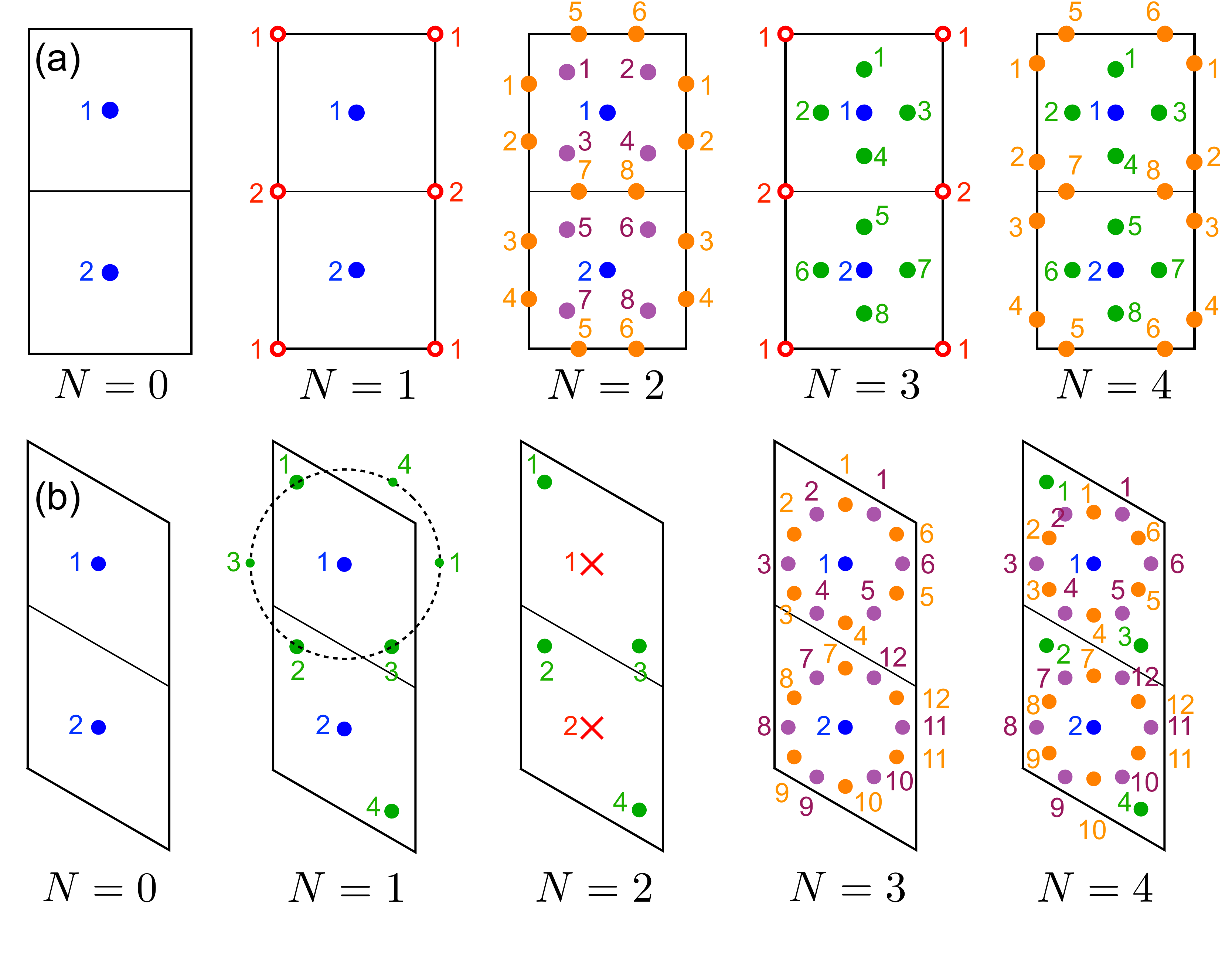}
	\caption{
    Counting of the gap closure points in the magnetic Brillouin zone for the transition lines in the vicinity of the Landau-level line $\mu = E_N$, where $N=0$--$4$. The numbers of the same color enumerate distinct gap-closure points related by the vortex-lattice symmetries and corresponding to the line of the same color in Fig~\ref{fig:tr-phase-diagram} and \ref{fig:sq-phase-diagram}.
    The superconducting vortex is (a) square; (b) triangular.
    \label{fig:gap-closures-numbered}}
\end{figure}
\end{widetext}

\bibliography{IQHE-SC}

\end{document}